# New Framework of Getis-Ord's Indexes Associating Spatial Autocorrelation with Interaction

Yanguang Chen

(Department of Geography, College of Urban and Environmental Sciences, Peking University, 100871, Beijing, China. Email: chenyg@pku.edu.cn.)

**Abstract:** Spatial autocorrelation and spatial interaction are two important analytical processes for geographical analyses. However, the internal relations between the two types of models have not been brought to light. This paper is devoted to integrating spatial autocorrelation analysis and spatial interaction analysis into a logic framework by means of Getis-Ord's indexes. Based on mathematical derivation and transform, the spatial autocorrelation measurements of Getis-Ord's indexes are reconstructed in a new and simple form. A finding is that the local Getis-Ord's indexes of spatial autocorrelation are equivalent to the rescaled potential energy indexes of spatial interaction theory based on power-law distance decay. The normalized scatterplot is introduced into the spatial analysis based on Getis-Ord's indexes, and the potential energy indexes are proposed as a complementary measurement. The global Getis-Ord's index proved to be the weighted sum of the potential energy indexes and the direct sum of total potential energy. The empirical analysis of the system of Chinese cities are taken as an example to illustrate the effect of the improved methods and measurements. The mathematical framework newly derived from Getis-Ord's work is helpful for further developing the methodology of geographical spatial modeling and quantitative analysis.

**Key words:** Spatial Autocorrelation; Spatial Interaction; Spatial modeling; Getis-Ord's *G*; Potential energy; Gravity model

## 1 Introduction

Spatial autocorrelation and spatial interaction models represent two theoretical cornerstones and classic contents of geographical analyses. Spatial autocorrelation is based on the concept of correlation coefficient, and the main measurements include Moran's index (Moran, 1948), Geary's coefficient (Geary, 1954), and Getis-Ord's indexes (Getis and Ord, 1992; Ord and Getis, 1995). Spatial interaction is based on the gravity concept, and the chief models and methods including gravity model (Fotheringham and O'Kelly, 1989; Haggett *et al*, 1977; Haynes and Fotheringham, 1984), potential energy formulae (Stewart, 1942; Stewart, 1948), and entropy-maximizing model family (Wilson, 1968; Wilson, 1970; Wilson, 2000). However, the mathematical links between spatial autocorrelation and spatial interaction have not been revealed at present. In fact, there are significant similarities and differences between the two methods. The similarities between spatial autocorrelation and interaction are as follows. First, both of them are based on size measurements and distance decay effect. Second, both of them can be used to describe strength patterns of spatial association between different geographical elements. The principal difference between the two methods rests with the correlation properties. Spatial autocorrelation is focused on the intra-correlation or self-correlation of a group of elements, while the spatial interaction is focused on the inter-correlation or cross-correlation between many different elements, especially two elements. Sometimes, if we examine the same elements in a geographical system by using the same size and distance measurements, auto-correlation and cross-correlation are often weaved into one another. Thus, spatial autocorrelation analysis may be combined with spatial interaction modeling. If so, we can find a new way of spatial analysis for characterizing geographical patterns and processes.

In a sense, spatial autocorrelation analyses are more widely made than spatial interaction analyses in scientific studies. The former is a theory of spatial statistics, while the latter is a geographical theoretical model. The methods of spatial autocorrelation have been developing (Anselin, 2019; Bivand, 2009; Carrijo and da Silva, 2017; Chen, 2013; Griffith, 2003; Haining, 2009; Lee and Li, 2017; Li *et al*, 2007; Liu *et al*, 2015; Ord and Getis, 1995; Sokal and Oden, 1978; Tiefelsdorf, 2002). The statistics of spatial autocorrelation such as Moran's $I$ and Ripley's $K$ has been applied to spatial association processes in various fields, for example, man-land relationships (Beck and Sieber, 2010), human diseases (Lai and Law, 2015; Melo *et al*, 2017; Ng *et al*, 2012; Rogerson, 2015; Wang *et al*, 2014), animal disease transmission (de la Cruz *et al*, 2014), human fertility and mortality (Kumar *et al*, 2012; Weeks *et al*, 2004), human genome (Koester *et al*, 2012), spatial pattern of urbanization (Scheuer *et al*, 2016), ecological patterns (Braun *et al*, 2012; Oreska *et al*, 2017; Wilson *et al*, 2014), maritime anomaly detection (Roberts, 2019), and spatial sampling and data analysis (Deblauwe *et* *al*, 2012; Jung *et al*, 2019; Mateo-Tomás and Olea, 2010; Wang *et al*, 2012; Westerholt *et al*, 2016). In contrast, spatial interaction analysis is mainly confined to geographical research. A discovery will be made in this work that the Getis-Ord's indexes can be used to connect spatial autocorrelation and spatial interaction based on the power-law decay. If we can express the inherent correlation between them by mathematical equations, we will be able to advance the methodology of spatial analysis. This paper is devoted to reconstructing the mathematical expressions of Getis-Ord's indexes and thus integrating the spatial interaction into spatial autocorrelation analysis using Getis-Ord's indexes. Solving this problem results in a series of improvements to the models and measurements based on the Getis-Ord's indexes. The rest parts are organized as follows. First, a new mathematical framework of spatial autocorrelation based on Getis-Ord's indexes are proposed, and a scatterplot is introduced into the new framework to visualize the analytical process. Then, the local Getis-Ord's indexes based on the power-law distance decay are proved to be the rescaled potential energy indexes, and the global Getis-Ord's index proved to be the weighted sum of the local indexes. Finally, the system of the main Chinese cities are taken as an example to illustrate how to use the new analytical framework of spatial autocorrelation process.

## 2 Theoretical results

### 2.1 Reconstructing formulae of Getis-Ord's indexes

In spatial autocorrelation analysis, Getis-Ord's indexes are important complement to Moran's indexes and Geary's coefficients. Using Getis-Ord's indexes, we can reveal the inherent relationship between spatial autocorrelation and spatial interaction. Firstly, the mathematical expression of Getis-Ord's indexes should be reconstructed in a new form. Then, we can reveal the mathematical relationships between Getis-Ord's indexes and potential indexes. Suppose that there are $n$ geographical elements (e.g., cities) in a regional system (e.g., a network of cities) which can be measured by a size variable $x$ (e.g., city population). A vector of the element sizes is as follows

$$\mathbf{x} = \begin{bmatrix} x_1 & x_2 & \cdots & x_n \end{bmatrix}^{\mathrm{T}}, \tag{1}$$

where $x_i$ is the size measurement of the $i$th element ($i$=1,2,…,$n$). The sum of $x_i$ is as below:

$$S = \sum_{i=1}^{n} x_i \, . \tag{2}$$

The unitized vector of **x** can be given by $\mathbf{y}=\mathbf{x}/S=[y_1, y_2, \ldots, y_n]^{\mathrm{T}}$, in which the $i$th entry is

$$y_i = \frac{x_i}{S} = x_i / \sum_{i=1}^{n} x_i = \frac{x_i}{n\bar{x}}, \tag{3}$$

in which $\bar{x}$ denotes the mean of $x_i$. The unitization processing depends on the mean of size variable, and average value represents the characteristic length of a sample. The concept of unitization based on sum is often confused with the notion of normalization based on range in literature. The variable $y$ meets the condition of unitization such as

$$\sum_{i=1}^{n} y_i = \sum_{i=1}^{n} (x_i / \sum_{i=1}^{n} x_i) = \frac{1}{S} \sum_{i=1}^{n} x_i = 1. \tag{4}$$

Thus, Getis-Ord's index $G$ can be re-expressed in a simple way by means of the unitized variable. Based on a spatial contiguity matrix (SCM), we can construct a spatial weight matrix (SWM). Suppose that there is an $n$-by-$n$ *unitized spatial weights matrix* (USWM) such as

$$\mathbf{W} = \left[ w_{ij} \right]_{n \times n}, \tag{5}$$

where $i, j=1,2,\ldots,n$. The three properties of the matrix are as follows: (1) Symmetry, i.e., $w_{ij}=w_{ji}$; (2) Zero diagonal elements, namely, $|w_{ii}|=0$; (3) Unitization condition, that is

$$\sum_{i=1}^{n} \sum_{j=1}^{n} w_{ij} = 1. \tag{6}$$

Thus the global Getis-Ord's index $G$ can be expressed in a quasi-quadratic form as follows

$$G = \mathbf{y}^{\mathrm{T}} \mathbf{W} \mathbf{y}, \tag{7}$$

which is simple and more convenient than the conventional expression of Getis-Ord's index. In fact, $G$ is not a really a quadratic form because **W** is not a positive definite matrix. Expanding equation (7) yields the original formula of Getis-Ord's index (Getis and Ord, 1992; Ord and Getis, 1995)

$$G = \sum_{i=1}^{n} \sum_{j=1}^{n} w_{ij} y_i y_j = \frac{\sum_{i=1}^{n} \sum_{j=1}^{n} w_{ij} x_i x_j}{\sum_{i=1}^{n} \sum_{j=1}^{n} x_i x_j}, \tag{8}$$

where $w_{ij}$ denotes the elements of a spatial weight matrix, **W** (Chen, 2013; Chen, 2015a). Equation (8) is the common mathematical expression of the global Getis-Ord's index. The local Getis's $G$ can be re-written as

$$\mathbf{G} = \mathbf{W} \mathbf{y}, \tag{9}$$

where $\mathbf{G}=[G_1, G_2,\ldots, G_n]^{\mathrm{T}}$. Accordingly, the expanded form is

$$G_i = \sum_{j=1}^{n} w_{ij}(x_j / \sum_{j=1}^{n} x_j) = \sum_{j=1}^{n} w_{ij} y_j \ , \tag{10}$$

which represents an important measurement of local spatial autocorrelation.

Now, we can investigate the association of spatial autocorrelation with spatial interaction. In fact, if we use the reciprocals of distances between geographical elements (locations) to construct a spatial contiguity matrix, equation (10) proved to be equivalent to the formula of potential energy. Proposed by Stewart (1948; 1950a, 1950b), potential energy is a useful measurement in urban geography (Zhou, 1995). In fact, the local Getis's *G* reflects a kind of normalized potential energy, and this will be demonstrated next. A normalized potential energy can be defined as follows

$$E_i = (x_i / \sum_{i=1}^{n} x_i) \sum_{j=1}^{n} (w_{ij} x_j / \sum_{j=1}^{n} x_j) = y_i \sum_{j=1}^{n} w_{ij} y_j \ , \tag{11}$$

which bears an analogy with local Moran's index in form. It can be termed the *Local Indicators of Spatial Interaction* (LISI), which bears an analogy with the local indicators of spatial association (LISA) (Anselin, 1995; Anselin, 1996). The *G* value is a relative measurement, while the *E* value is an absolute measurement for spatial association. It can be proved that

$$G = \sum_{i=1}^{n} E_i = \sum_{i=1}^{n} y_i \sum_{j=1}^{n} w_{ij} y_j = \sum_{i=1}^{n} \sum_{j=1}^{n} w_{ij} y_i y_j \ , \tag{12}$$

which indicates that the global Getis-Ord's index *G* equals the sum of the total potential energy $E_i$.

Scientific description based on mathematical theory is to utilize characteristic scales, which can be represented by eigenvalues in linear algebra. The theoretical eigen equation of Getis's index can be derived from the abovementioned definitions. Equation (7) multiplied left by vector **y** on both sides of the equal sign yields

$$\mathbf{M}^{*}\mathbf{y} = \mathbf{y}\mathbf{y}^{\mathrm{T}}\mathbf{W}\mathbf{y} = G\mathbf{y} , \tag{13}$$

where

$$\mathbf{M}^{*} = \mathbf{y}\mathbf{y}^{\mathrm{T}}\mathbf{W} \tag{14}$$

can be termed the Ideal Spatial Correlation Matrix (ISCM) in a theoretical sense. ISCM is the outer product correlation matrix (OPCM). In equation (13), **y** is the eigenvector (characteristic vector) of $\mathbf{M}^{*}$ and Getis-Ord's index *G* is just the corresponding maximum eigenvalue (characteristic root).

Expanding equation (13) yields

$$\begin{bmatrix} y_1 \\ y_2 \\ \vdots \\ y_n \end{bmatrix} \begin{bmatrix} y_1 & y_2 & \cdots & y_n \end{bmatrix} \begin{bmatrix} w_{11} & w_{12} & \cdots & w_{1n} \\ w_{21} & w_{22} & \cdots & w_{2n} \\ \vdots & \vdots & \ddots & \vdots \\ w_{n1} & w_{n2} & \cdots & w_{nn} \end{bmatrix} = \begin{bmatrix} y_1\sum_{j=1}^{n} w_{1j}y_j & y_1\sum_{j=1}^{n} w_{2j}y_j & \cdots & y_1\sum_{j=1}^{n} w_{nj}y_j \\ y_2\sum_{j=1}^{n} w_{1j}y_j & y_2\sum_{j=1}^{n} w_{2j}y_j & \cdots & y_2\sum_{j=1}^{n} w_{nj}y_j \\ \vdots & \vdots & \ddots & \vdots \\ y_n\sum_{j=1}^{n} w_{1j}y_j & y_n\sum_{j=1}^{n} w_{2j}y_j & \cdots & y_n\sum_{j=1}^{n} w_{nj}y_j \end{bmatrix} \quad (15)$$

which is important for the autocorrelation analysis based on Getis-Ord's indexes. Comparing equation (15) with equation (11) shows that the elements in the diagonal of $\mathbf{M}^*$ give the normalized potential energy of a geographical system. The trace of $\mathbf{M}^*$ is equal to the global Getis-Ord's index, $G$. The sum of each volume of $\mathbf{M}^*$ yields the local Getis' $G$, that is

$$E_k = \sum_{i=1}^{n} y_i \sum_{j=1}^{n} w_{kj} y_j \,, \quad (16)$$

where $i, j, k = 1,2,\ldots,n$. Please note that equation (16) is different from equation (12). The sum of each row of $\mathbf{M}^*$ gives the product of $y_i$ and the sum of $G_i$, namely,

$$y_i \sum_{k=1}^{n} \sum_{j=1}^{n} w_{kj} y_j = y_i \sum_{i=1}^{n} G_i \,, \quad (17)$$

which implies

$$\sum_{i=1}^{n} G_i = \sum_{k=1}^{n} \sum_{j=1}^{n} w_{kj} y_j = \sum_{i=1}^{n} \sum_{j=1}^{n} w_{ij} y_j \,, \quad (18)$$

where $i, j, k = 1,2,\ldots,n$. Equations (16), (17), and (18) can be verified by a simple example. This suggests that we can calculate the normalized potential energy, potential energy indexes, global Getis-Ord's index, and local Getis-Ord's indexes by means of the matrix $\mathbf{M}^*$.

## 2.2 Actual spatial correlation matrix

The practical spatial correlation matrix is different from the ideal spatial correlation matrix. In empirical studies, the outer product $\mathbf{y}\mathbf{y}^{\mathrm{T}}$ in equation (13) can be substituted with the inner product $\mathbf{y}^{\mathrm{T}}\mathbf{y}$. In fact, the result of $\mathbf{y}^{\mathrm{T}}\mathbf{y}$ is a constant. So we have

$$\mathbf{y}\mathbf{y}^{\mathrm{T}}\mathbf{y} = \mathbf{y}^{\mathrm{T}}\mathbf{y}\mathbf{y} = \lambda\mathbf{y} \,, \quad (19)$$

which suggests that the parameter $\lambda = \mathbf{y}^{\mathrm{T}}\mathbf{y}$ is the maximum eigenvalue of the outer product matrix

$\mathbf{yy}^{\mathbf{T}}$, and the unitized size vector $\mathbf{y}$ is the corresponding eigenvector. Developing equation (19) yields

$$\begin{bmatrix} y_1 \\ y_2 \\ \vdots \\ y_n \end{bmatrix} \begin{bmatrix} y_1 & y_2 & \cdots & y_n \end{bmatrix} \begin{bmatrix} y_1 \\ y_2 \\ \vdots \\ y_n \end{bmatrix} = \begin{bmatrix} y_1 \sum_{i=1}^{n} y_i^2 \\ y_2 \sum_{i=1}^{n} y_i^2 \\ \vdots \\ y_n \sum_{i=1}^{n} y_i^2 \end{bmatrix} = \lambda \begin{bmatrix} y_1 \\ y_2 \\ \vdots \\ y_n \end{bmatrix}. \tag{20}$$

Further, it can be shown that $\lambda=\mathbf{y}^{\mathbf{T}}\mathbf{y}$ is the maximum eigenvalue of $\mathbf{yy}^{\mathbf{T}}$. For a square matrix, the trace of $\mathbf{yy}^{\mathbf{T}}$ is

$$\mathrm{T_r}(\mathbf{yy}^{\mathbf{T}}) = \sum_{i=1}^{n} y_i^2 = \lambda = \lambda_1 + \lambda_2 + \cdots + \lambda_n , \tag{21}$$

where $\mathrm{T_r}$ refers to "finding the trace (of $\mathbf{yy}^{\mathbf{T}}$)". If $\lambda_1=\lambda_{\max}=\mathbf{y}^{\mathbf{T}}\mathbf{y}$, then we will have

$$\lambda = \begin{cases} \mathbf{y}^{\mathbf{T}}\mathbf{y}, & \lambda = \lambda_{\max} \\ 0, & \lambda \neq \lambda_{\max} \end{cases}. \tag{22}$$

For arbitrary $\lambda$, the extended form of $\mathbf{yy}^{\mathbf{T}}$ is as below:

$$\mathbf{yy}^{\mathbf{T}} = \begin{bmatrix} y_1 \\ y_2 \\ \vdots \\ y_n \end{bmatrix} \begin{bmatrix} y_1 & y_2 & \cdots & y_n \end{bmatrix} = \begin{bmatrix} y_1 y_1 & y_1 y_2 & \cdots & y_1 y_n \\ y_2 y_1 & y_2 y_2 & \cdots & y_2 y_n \\ \vdots & \vdots & \ddots & \vdots \\ y_n y_1 & y_n y_1 & \cdots & y_n y_n \end{bmatrix}. \tag{23}$$

According to the Cayley-Hamilton theorem, the eigenvalues of any $n$-by-$n$ matrix are identical to the characteristic roots of the polynomial equation. The characteristic polynomial results from the determinant of the matrix $\mathbf{yy}^{\mathbf{T}}$, that is

$$\lambda\mathbf{E} - \mathbf{yy}^{\mathbf{T}} = \begin{vmatrix} \lambda - y_1 y_1 & -y_1 y_2 & \cdots & -y_1 y_n \\ -y_2 y_1 & \lambda - y_2 y_2 & \cdots & -y_2 y_n \\ \vdots & \vdots & \ddots & \vdots \\ -y_n y_1 & -y_n y_1 & \cdots & \lambda - y_n y_n \end{vmatrix} = 0 , \tag{24}$$

where $\mathbf{E}$ denotes the identity/unit matrix. Finding the characteristic roots of equation (24) yields $\lambda_1=\lambda_{\max}=\mathbf{y}^{\mathrm{T}}\mathbf{y}=y_1^2+y_2^2+\ldots+y_n^2$ and $\lambda_2=\lambda_3=\ldots=\lambda_n=0$.

Now, a practical autocorrelation expression based on the global Getis-Ord's index can be given by matrixes and vectors. Substituting the maximum eigenvalue $\lambda$ for the corresponding matrix $\mathbf{yy}^{\mathbf{T}}$ in equation (13) products a new mathematical relation. The precondition that equation (7) comes into existence is

$$\lambda \mathbf{W}\mathbf{y} = G\mathbf{y}. \tag{25}$$

In fact, equation (25) is left multiplied by $\mathbf{y}^{\mathrm{T}}$ yields equation (7). This implies that we can derive equation (7) from equation (25). Obviously, Getis-Ord's index is the maximum eigenvalue of the weight matrix $\lambda\mathbf{W}$, and $\mathbf{y}$ is the corresponding eigenvector, which can be normalized as $\mathbf{y}/\sqrt{\lambda}$. Equation (25) can be re-expressed as a matrix scaling relation such as

$$\mathbf{M}\mathbf{y} = \lambda \mathbf{W}\mathbf{y} = \mathbf{y}^{\mathrm{T}}\mathbf{y}\mathbf{W}\mathbf{y} = G\mathbf{y}, \tag{26}$$

where

$$\mathbf{M} = \lambda \mathbf{W} = \mathbf{y}^{\mathrm{T}}\mathbf{y}\mathbf{W}. \tag{27}$$

In this equation, **M** can be termed the Real Spatial Correlation Matrix (RSCM) in the sense of application. RSCM is the inner product correlation matrix (IPCM). The trace of the matrix $\lambda\mathbf{W}$ is the eigenvalue with the minimum absolute value, i.e. $T_r(\lambda\mathbf{W})=0$. Normalizing the eigenvector yields

$$\mathbf{y}^{\mathbf{o}} = \frac{\mathbf{y}}{\sqrt{\|\mathbf{y}\|}} = \frac{\mathbf{y}}{\sqrt{\lambda}}. \tag{28}$$

If we use the mathematical software such as Matlab to calculate the eigenveactor of $\mathbf{y}\mathbf{y}^{\mathrm{T}}\mathbf{W}$ or $\lambda\mathbf{W}$, the result will be $\mathbf{y}^{\circ}$ rather than $\mathbf{y}$. Comparing equation (25) with equation (13) shows

$$\mathbf{y}\mathbf{y}^{\mathrm{T}}\mathbf{W}\mathbf{y} = \lambda \mathbf{W}\mathbf{y}. \tag{29}$$

This indicates that the eigenvector $\mathbf{G}=\mathbf{W}\mathbf{y}$ is still the eigenvector of the outer product matrix $\mathbf{y}\mathbf{y}^{\mathrm{T}}$, and the corresponding eigenvalue is $\lambda= \mathbf{y}^{\mathrm{T}}\mathbf{y}$. Substituting equation (9) into equation (29) yields

$$\mathbf{y}\mathbf{y}^{\mathrm{T}}\mathbf{G} = \lambda \mathbf{G}, \tag{30}$$

which suggests that the vector of local Getis-Ord's index is the eigenvector of $\mathbf{y}\mathbf{y}^{\mathrm{T}}$ corresponding to the eigenvalue $\lambda$. Thus we have

$$(\lambda \mathbf{E} - \mathbf{y}\mathbf{y}^{\mathrm{T}})\mathbf{W}\mathbf{y} = (\lambda \mathbf{W} - \mathbf{y}\mathbf{y}^{\mathrm{T}}\mathbf{W})\mathbf{y} = 0, \tag{31}$$

in which 0 refers to the zero/null vector. However, equations (29) and (31) cannot occur unless the spatial contiguity matrix is a unit matrix. In other words, the vector **G** is not really an eigenvector of $\mathbf{y}\mathbf{y}^{\mathrm{T}}$. In empirical analysis, the null vector should be replaced by a residual vector. An approximation relation is as follows

$$\mathbf{M}\mathbf{y} = \lambda \mathbf{W}\mathbf{y} \to \mathbf{y}\mathbf{y}^{\mathrm{T}}\mathbf{W}\mathbf{y} = \mathbf{M}^{*}\mathbf{y}, \tag{32}$$

where the arrow "→" denotes "infinitely approach to" or "be theoretically equal to". There are always errors between the inner product correlation matrix $\mathbf{M}=\mathbf{y}^{\mathrm{T}}\mathbf{y}\mathbf{W}$ and the outer product correlation matrix $\mathbf{M}^{*}=\mathbf{y}\mathbf{y}^{\mathrm{T}}\mathbf{W}$. Based on the error vector, we can define an index to measure the degree of spatial autocorrelation. The stronger the spatial autocorrelation is, the closer the vector $\mathbf{My}$ will be to the vector $\mathbf{M}^{*}\mathbf{y}$. A finding is that, according to the equations (13) and (26), the global Getis-Ord's index proved to be the eigenvalue of spatial correlation matrixes. As indicated above, an eigenvalue of a matrix is the characteristic root of the corresponding multinomial of the determinant of the matrix. It represents a characteristic length of spatial analysis. This suggests that, like Moran's $I$, Getis-Ord's $G$ is also a characteristic parameter of geographical spatial modeling.

## 2.3 Getis-Ord's scatterplot

The spatial analytical process based on Getis-Ord's index can be visualized by scatter plots. In order to find new approaches to evaluating Getis-Ord's indexes and introducing Getis-Ord's scatterplot into spatial autocorrelation analysis, two vectors based on spatial correlation matrixes should be defined. One is the outer product vector as below

$$\mathbf{f}^{*}=\mathbf{M}^{*}\mathbf{y}=\mathbf{y}\mathbf{y}^{\mathrm{T}}\mathbf{W}\mathbf{y}=G\mathbf{y}, \tag{33}$$

which is based on equation (13). The other is the inner product vector as follows

$$\mathbf{f}=\mathbf{My}=\mathbf{y}^{\mathrm{T}}\mathbf{y}\mathbf{W}\mathbf{y}=G\mathbf{y}, \tag{34}$$

which is based on equation (26). The relationship between $\mathbf{y}$ and $\mathbf{f}^{*}$ suggests the theoretical autocorrelation trend line, and the dataset of $\mathbf{y}$ and $\mathbf{f}$, indicates the scatter points of actual autocorrelation pattern. The residuals of spatial autocorrelation can be defined as

$$\mathbf{e}_{\mathbf{f}}=\mathbf{f}-\mathbf{f}^{*}=\mathbf{My}-\mathbf{M}^{*}\mathbf{y}=(\lambda\mathbf{E}-\mathbf{y}\mathbf{y}^{\mathrm{T}})\mathbf{W}\mathbf{y}, \tag{35}$$

where $e_f$ refers to the errors of the Getis-Ord's spatial autocorrelation. The squared sum of the residuals $S_f$ is

$$S_{f}=\mathbf{e}_{\mathbf{f}}^{\mathrm{T}}\mathbf{e}_{\mathbf{f}}=\mathbf{y}^{\mathrm{T}}\mathbf{W}(\lambda\mathbf{E}-\mathbf{y}\mathbf{y}^{\mathrm{T}})(\lambda\mathbf{E}-\mathbf{y}\mathbf{y}^{\mathrm{T}})\mathbf{W}\mathbf{y}\to 0. \tag{36}$$

The value of $e_f$ fluctuates around 0; therefore, the $S_f$ value approaches zero.

By analogy with Moran's scatterplot, we can employ scatter point graphs to make local spatial autocorrelation analysis based on Getis-Ord's indexes. If the unitary vector $\mathbf{y}$ represents the $x$-axis, and the corresponding vector $\lambda\mathbf{W}\mathbf{y}$ represents the $y$-axis, a Getis-Ord's scatterplot will be generated.

Further, a "trend line" can be added to the plot: the $x$-axis is still the unitary vector, $\mathbf{y}$, but the $y$-axis is $\mathbf{yy^TWy}$. In other words, the relationship between $\mathbf{y}$ and $\lambda\mathbf{Wy}$ forms the scatter points, while the relationship between $\mathbf{y}$ and $\mathbf{yy^TWy}$ makes the trend line. Differing from Moran's index which comes between -1 and 1, Getis-Ord's index ranges from 0 to 1. That is to say, $G\geq0$. As a result, the trend line based on $\mathbf{yy^TWy}$ does not always match the scatter points based on $\lambda\mathbf{Wy}$. In fact, for the positive spatial autocorrelation (Moran's $I>0$), a Getis-Ord's trend line is consistent with its scatter points; however, for the negative spatial autocorrelation (Moran's $I<0$), a Getis-Ord's trend line is inconsistent with its scatter points. In many cases, a trend line of Getis-Ord's scatter plot serves for a dividing line, and the data points fall into two categories. By means of the scatter points and trend line, we can divide the geographical elements into two groups.

# 3 Discussion

## 3.1 Association of autocorrelation with interaction

So far, a series of improvement and development of the spatial autocorrelation analysis based on Getis-Ord's indexes have been fulfilled. Using the improved expressions of Getis-Ord's indexes, we can associate spatial autocorrelation analysis with spatial interaction analysis. The main findings and innovations of this work are as follows. First, the computational formulae of Getis-Ord's indexes are simplified and normalized. Unitizing size vector and spatial weight matrix, we can express Getis-Ord's index in the simpler way so that the calculations become easier. Second, a scatter plot can be introduced into the analytical process. By analogy with Moran's scatter plot, we can draw a scatter plot for Getis-Ord's autocorrelation analysis. Using the scatter plot, we can visualize the spatial patterns and divide geographical elements into several groups. Third, Getis-Ord's index proved to be an eigenvalue of a spatial correlation matrix. This suggests that Getis-Ord's index is actually a characteristic length of spatial autocorrelation. Fourth, if we use the reciprocals of geographical distances to define spatial contiguity, Getis-Ord's index is demonstrated to be equivalent to potential energy. Suppose that spatial contiguity matrix is generated using power-law decay and the distance decay exponent equals 1. Getis-Ord's index can be converted into local potential energy. Thus, spatial autocorrelation is mathematically associated with spatial interaction.

The precondition of the abovementioned innovations is reconstruction of Getis-Ord's index formula with matrixes and vectors. It is easy to prove the following relation:

$$\sum_{i=1}^{n}\sum_{j=1}^{n}x_i x_j = \sum_{i=1}^{n}x_i \sum_{j=1}^{n}x_j , \tag{37}$$

where

$$\sum_{i=1}^{n}x_i = \sum_{j=1}^{n}x_j = const , \tag{38}$$

in which *const* denotes a constant. Thus, re-expressing equation (8) yields

$$G = \frac{\sum_{i=1}^{n}\sum_{j=1}^{n}w_{ij}x_i x_j}{\sum_{i=1}^{n}x_i \sum_{j=1}^{n}x_j} = \sum_{i=1}^{n}\sum_{j=1}^{n}w_{ij}\left(\frac{x_i}{\sum_{i=1}^{n}x_i}\frac{x_j}{\sum_{j=1}^{n}x_j}\right) = \sum_{i=1}^{n}\sum_{j=1}^{n}w_{ij}y_i y_j , \tag{39}$$

which is equivalent to equation (7). The relation between the global Getis-Ord's index and the local Getis-Ord's index is

$$G = \sum_{i=1}^{n}\frac{x_i}{\sum_{i=1}^{n}x_i}\sum_{j=1}^{n}w_{ij}\left(\frac{x_j}{\sum_{j=1}^{n}x_j}\right) = \sum_{i=1}^{n}y_i \sum_{j=1}^{n}w_{ij}y_j = \sum_{i=1}^{n}y_i G_i , \tag{40}$$

in which $G_i$ is defined by equation (9). It is obvious that equation (40) is equivalent to equations (12) and (16). This suggests that the global Getis-Ord's index is the weighted sum of local Getis-Ord's index based on the unitized size vector.

By comparison, the relationships and differences between Getis-Ord's indexes, Moran's indexes, and potential energy indexes can be made clearer. Getis-Ord's indexes are different from Moran's indexes. Getis and Ord (1992) proposed the indexes to make up the deficiencies of Moran's indexes. However, there is an analogy between Getis-Ord's *G* and Moran's *I*. The similarities are as follows. First, the method of improving the mathematical expressions of Getis-Ord's index is similar to that of improving the mathematical expressions of Moran's index. Second, both Moran's *I* and Getis-Ord's *G* proved to be the eigenvalues of spatial correlation matrixes. Third, both the two computational processes depend on the variable transformation based on average values. The eigenvalues represent the characteristic length of spatial correlation, while average values represent the characteristic length of size samples. A comparison between the two measurements is drawn and tabulated as follows (Table 1). Apparently, both the new forms of the Getis-Ord's indexes and Moran's indexes are based on unitized spatial contiguity matrix, **W**. But the size vector is different in form. The Moran's indexes are based on standardized size vector, while the corresponding Getis-Ord's indexes is based on unitized size vector. So, Moran's index $I$ comes between -1 and 1 ($-1 \le I \le 1$), while Getis-Ord's index $G$ varies from 0 to 1 ($0 \le G \le 1$).

**Table 1 A comparison of form and structure between Moran's index, *I*, and Getis-Ord's index, *G***

| **Parameter** | **Formula** | | **Definition of variable** |
|---|---|---|---|
| | Global index | Local index | |
| **Moran's index, *I*** | $I = \mathbf{z}^{\mathrm{T}}\mathbf{W}\mathbf{z}$ | $I_i = z_i \sum_{j=1}^{n} w_{ij} z_j$ | $z_i = (x_i - \bar{x}) / s$ |
| **Getis-Ord's index, *G*** | $G = \mathbf{y}^{\mathrm{T}}\mathbf{W}\mathbf{y}$ | $G_i = \sum_{j=1}^{n} w_{ij} y_j$ | $y_i = x_i / \sum_{i=1}^{n} x_i = x_i / (n\bar{x})$ |

Next, let's investigate the relationship between Getis-Ord's indexes for spatial autocorrelation and the potential energy indexes for spatial interaction. The classical gravity model of geographical spatial interaction is as below (Haggett *et al*, 1977):

$$I_{ij} = K \frac{x_i x_j}{r_{ij}^b}, \tag{41}$$

where $x_i$ and $x_j$ are two size measures (e.g., city population), $r_{ij}$ is the distance between the $i$ location and the $j$ location, $I_{ij}$ denotes the attraction force between $x_i$ and $x_j$, the parameter $K$ refers to the gravity coefficient, and $b$ to the distance decay exponent ($b$>1). The distance exponent proved to be a kind of fractal dimension (Chen, 2015b). Thus the mutual energy between the $i$ location and the $j$ location can be defined as (Stewart, 1948; Stewart, 1950; Stewart and Warntz, 1958)

$$I_{ij} r_{ij} = K \frac{x_i x_j}{r_{ij}^{b-1}}. \tag{42}$$

Thus, the gravitational potential can be defined as $s_j = I_{ij} r_{ij} / x_i$ (Stewart and Warntz, 1958). The total mutual energy (TME) between the $i$ location and other locations can be given by

$$E_i = \sum_{j=1}^{n} I_{ij} r_{ij} = K x_i \sum_{j=1}^{n-1} \frac{x_j}{r_{ij}^{b-1}} = K x_i \sum_{j=1}^{n-1} \frac{x_j}{r_{ij}^{q}}. \tag{43}$$

where $q=b-1$ denotes distance scaling exponent. The value of $E_i$ reflects the influence power of an element at the $i$th location in a regional network. Accordingly, the potential energy index (PEI) indicating the total gravitational potential of the $i$ location in a geographical system can be defined as (Zhou, 1995)

$$V_i = \frac{E_i}{x_i} = K\sum_{j=1}^{n-1}\frac{x_j}{r_{ij}^{b-1}} = K\sum_{j=1}^{n-1}\frac{x_j}{r_{ij}^{q}}, \tag{44}$$

which reflects the traffic accessibility of location $i$. Without loss of generality, let $K$=1 and $b$=2, then we have $q$=1. Suppose that the spatial proximity function (SPF) is $v_{ij}=1/r_{ij}$ and $x_i$ and $x_j$ are replaced by $y_i$ and $y_j$. Unitizing the spatial contiguity matrix, we can convert equation (44) into equation (10), and transform equation (43) into equation (11). This suggests that Getis-Ord's index is actually normalized potential energy, and spatial autocorrelation analysis and spatial interaction modeling reach the same goal by different routes.

### 3.2 Equivalence of Getis-Ord's *G* to potential energy

In order to further reveal the association of spatial autocorrelation with spatial interaction, the clearer and exacter relation between Getis-Ord's indexes and potential energy should be shown. Now, let's change an angle of view to examine them. In fact, by rescaling potential energy of geographical elements, we can obtained local Getis-Ord's indexes. By the mathematical derivation, we can find practical links between the two approaches of spatial modeling. To make a spatial autocorrelation analysis, a spatial contiguity matrix must be created by applying a weight function to a spatial proximity matrix (Chen, 2012; Getis, 2009). For $n$ elements in a geographic system, a spatial contiguity matrix, $\mathbf{V}$, can be expressed as

$$\mathbf{V} = \left[v_{ij}\right]_{n\times n} = \begin{bmatrix} v_{11} & v_{12} & \cdots & v_{1n} \\ v_{21} & v_{22} & \cdots & v_{2n} \\ \vdots & \vdots & \ddots & \vdots \\ v_{n1} & v_{n2} & \cdots & v_{nn} \end{bmatrix}, \tag{45}$$

in which $v_{ij}$ is a measure used to reflect the contiguity relationships between location $i$ and location $j$ ($i$, $j$=1,2,…,$n$). If $i$=$j$ as given, then $v_{ii}$≡0. This indicates that the diagonal elements must be converted into zero. Thus a unitized spatial weights matrix, $\mathbf{W}$, can be given by

$$\mathbf{W} = \frac{\mathbf{V}}{V_0} = \begin{bmatrix} w_{11} & w_{12} & \cdots & w_{1n} \\ w_{21} & w_{22} & \cdots & w_{2n} \\ \vdots & \vdots & \ddots & \vdots \\ w_{n1} & w_{n2} & \ddots & w_{nn} \end{bmatrix}, \tag{46}$$

where

$$V_0 = \sum_{i=1}^{n}\sum_{j=1}^{n} v_{ij}, \quad w_{ij} = \frac{v_{ij}}{\sum_{i=1}^{n}\sum_{j=1}^{n} v_{ij}}, \quad \sum_{i=1}^{n}\sum_{j=1}^{n} w_{ij} = 1.$$

In above equations, the value $v_{ii}\equiv 0$ results in the value $w_{ii}\equiv 0$. Compared with spatial contiguity matrix **V**, the unitized spatial weights matrix **W** make the mathematical form of spatial autocorrelation become simple and graceful. If the spatial contiguity matrix is unitized by row, the result will violate the well-known distance axiom (Chen, 2016). There are three types of spatial weight function that can be used to construct spatial continuity matrix, that is, inverse power function, negative exponential function, and staircase functions (Chen, 2012). Among these weight functions, the inverse power function is the common one (Cliff and Ord, 1973). This function stemmed from the impedance function of the gravity model (Haggett *et al*, 1977). Generally speaking, the inverse power function is as below

$$v_{ij} = \begin{cases} r_{ij}^{-q}, & i \neq j \\ 0, & i = j \end{cases}, \tag{47}$$

where $r_{ij}$ refers to the distance between location $i$ and location $j$, and $q$ denotes the distance scaling exponent. Generally, we have $q$=1 for spatial autocorrelation (Cliff and Ord, 1981). A total quantity of spatial continuity can be defined as

$$S = \sum_{i=1}^{n}\sum_{j=1}^{n} r_{ij}^{-q}. \tag{48}$$

Then, we can rescale the spatial distances as follows

$$d_{ij} = (r_{ij}^{q} S)^{1/q}. \tag{49}$$

Based on the unitized size measure $y_j$ and rescaled distances $d_{ij}$, the potential energy is

$$V_i^* = \sum_{j=1}^{n} \frac{y_j}{d_{ij}^{q}}, \tag{50}$$

which can be regarded as rescaled potential energy. Based on the rescaled distances, the unitized weight is as below

$$w_{ij} = d_{ij}^{-q} = \frac{1}{r_{ij}^{q} S} = \frac{r_{ij}^{-q}}{\sum_{i=1}^{n}\sum_{j=1}^{n} r_{ij}^{-q}}. \tag{51}$$

Substituting equation (51) into equation (50) yields the normalized potential energy index

$$V_i^* = \sum_{j=1}^{n} \frac{y_j}{d_{ij}^q} = \sum_{j=1}^{n} w_{ij} y_j = G_i , \tag{52}$$

which suggests that the rescaled potential energy index $V_i^*$ equals local Getis-Ord's index $G_i$. Accordingly, the mutual energy index is $E_i^*=y_iV_i^*=y_iG_i$. That is to say, Getis-Ord's indexes for spatial autocorrelation are equivalent to the potential energy indexes for spatial interaction based on the gravity model under certain conditions.

This is a theoretical and methodological study for spatial autocorrelation and spatial interaction. Compared with pure autocorrelation measurements based on Getis-Ord's indexes, the new framework can yield more systematic outputs of calculations and analyses. The equivalence relationship between Getis-Ord's indexes and potential energy indexes is useful for spatial modeling. We can employ the gravity analysis of a regional network to estimate the distance scaling exponent value of spatial autocorrelation $q$. What is more, we can use spatial autocorrelation analysis to complement the spatial interaction analysis and *vice versa*. Getis-Ord's indexes are abstract and thus difficult to understand, but it is easy to understand the potential energy concept based on the gravity model. The chief shortcomings of this work are as follows. First, the method relies heavily on linear algebra theory. For the readers who are not familiar with linear algebra, especially matrix knowledge, it is hard to understand the methodology developed in this work. Second, the spatial autocorrelation and cross-correlation analyses are not integrated into framework. The spatial autocorrelation measures can be generalized to spatial cross-correlation measures (Chen, 2015a). Using total potential energy, we can associate spatial interaction with spatial autocorrelation and spatial cross-correlation. Due to the limited space, the problem remains to be solved in a companion paper.

# 4 Materials and Methods

## 4.1 Approaches to Getis-Ord's indexes

It is difficult for the learners of spatial autocorrelation and spatial interaction to compute Getis-Ord's index using the complex formulae. Students can calculate Getis-Ord's $G$ by means of the professional software such as ArcGIS. However, the computational process is a black box for them. If and only if a student knows how to fulfil a set of complete calculation steps of a measurement, he/she will really understand the principle of the mathematical method. Based on the new framework of Getis-Ord's spatial autocorrelation expressed by linear algebra, a number of approaches to computing global and local Getis-Ord's indexes are proposed in this section. Each approach has its own advantages and disadvantages (Table 2). Using the calculation results, we can make an analysis of spatial interaction with the potential energy values (Figure 1). Among these approaches, three ones bear analogy with those for Moran's index (Chen, 2013). In other words, all the approaches to calculating Moran's index can be employed to compute global Getis-Ord's index. The difference lies in the processing way of size measurements. However, for the local Getis-Ord's indexes, we should address them in the means differing from those for local Moran's indexes.

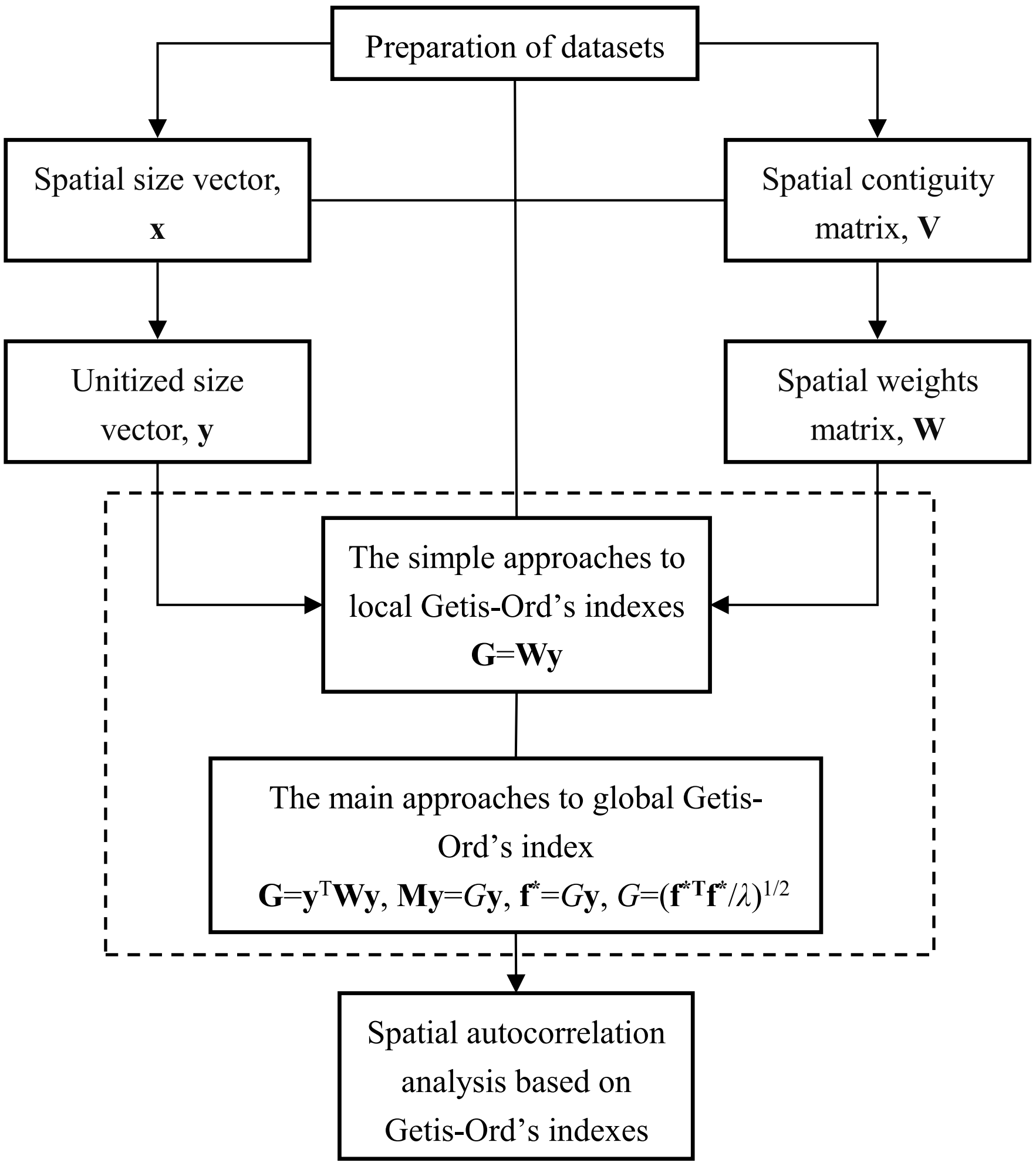

**Figure 1 A flow chart of data processing, parameter estimation, and autocorrelation analysis based on Getis-Ord's indexes**

[**Note**: The analytical process is similar to that based on Moran's index and Geary's coefficient. However, the measurements and conclusions are different. ]

The main approaches to computing the local Getis-Ord's indexes are as follows. **(1) Conventional formula method**. Using equation (10), we can calculate local Getis-Ord's indexes step by step. This is the traditional approach used in literature. **(2) Matrix manipulation method**. The sizes and weights must be unitized by equations (3) and (46). Then, in terms of equation (9), using the unitized weight matrix **W** to multiple left the unitized size vector **y** yields the vector of local Getis-Ord's indexes **G**. The process is very simple and can be carried out by MS Excel. **(3) Spatial correlation matrix method**. Suppose that we obtain the ideal spatial correlation matrix, $\mathbf{M}^*=\mathbf{y}\mathbf{y}^T\mathbf{W}$. According to equation (16), the sums of the columns of matrix $\mathbf{M}^*$ give the local Getis indexes. **(4) Potential energy method.** Local Getis-Ord's indexes are equal to the rescaled potential energy measurements. Using equation (3) to unitize size measurements, using equation (48) and (49) to rescale distance matrix, and using equation (52) to calculate the potential energy based on the special distance scaling exponent $q$=1, we can obtain the local Getis-Ord's indexes.

The approaches for calculating global Getis-Ord's index are more than seven ones, which are summarized as follows. **(1) Conventional formula method**. Using equation (8), we can compute the global Getis-Ord's index by the traditional method. **(2) Three-step calculation method.** This approach is very simple and the beginners of spatial autocorrelation analysis can master it easily. The three steps of calculating Getis-Ord's index are as follows. Step 1: unitize the size variable $x$. In other words, convert the initial variable $x$ based on equation (1) into the unitized variable in equation (3). Step 2: compute the unitized spatial weight matrix. The weights matrix is defined in equations (5) and (6) and can be calculated by equation (45) and (46). Step 3: calculate Getis-Ord's index. According to equation (7), the unitized spatial weight matrix is first left multiplied by the transposition of **y**, and then the vector $\mathbf{y}^T\mathbf{W}$ is right multiplied by **y**. The final product of the continued multiplication is the global Getis-Ord's index. **(3) Matrix scaling method.** This approach is to find the maximum characteristic value of the spatial correlation matrix. If we work out the maximum eigenvalue of the matrix $\mathbf{M}^*=\mathbf{y}\mathbf{y}^T\mathbf{W}$ or $\mathbf{M}=\lambda\mathbf{W}$ by using equation (13) or equation (26), we will gain the global Getis-Ord's index. **(4) Regression analysis method.** Based on equation (13) or equation (26), a linear regression analysis can be employed to evaluate Getis-Ord's $G$. The unitized vector **y** is treated as an independent variable (i.e., argument), and $\mathbf{f}^*=\mathbf{M}^*\mathbf{y}$ or $\mathbf{f}=\mathbf{M}\mathbf{y}$ as the corresponding dependent variable (response variable). If the constant term (intercept) is fixed to zero, the regression coefficient (slope) will be equal to the global Getis-Ord's index. **(5) Local** **weighting method**. After calculating the local Getis-Ord's indexes, we can figure out the global index using equation (40). The elements of the unitized size vector, **y**, can serve as weight numbers. The global Getis-Ord's index equals the weighted sum of the local indexes. **(6) Spatial correlation matrix method**. Using equation (16), we can generate the ISCM, $\mathbf{M}^*=\mathbf{y}\mathbf{y}^{\mathrm{T}}\mathbf{W}$. The trace, i.e., the sum of the diagonal elements of matrix $\mathbf{M}^*$, give the global Getis-Ord's index. **(7) Outer product sum method.** In terms of equation (4), the sum of **y**'s elements is 1. According to equation (33), we have

$$\sum_{i=1}^{n}(\mathbf{f}^*)_i = G\sum_{i=1}^{n}(\mathbf{y})_i = G\sum_{i=1}^{n} y_i = G\,. \tag{53}$$

Thus the value of Getis-Ord's index can be calculated using the elements in the vector $\mathbf{f}^*$, that is

$$G = \sum_{i=1}^{n} f_i^* = \sum_{i=1}^{n}(\mathbf{y}\mathbf{y}^{\mathrm{T}}\mathbf{W}\mathbf{y})_i\,, \tag{54}$$

which indicates an alternative approach to working out global Getis-Ord's index.

**Table 2 Comparison of the advantages and disadvantages of different approaches to global and local Getis-Ord's indexes**

| Level | Method | Simplicity | Result | Equation |
|---|---|---|---|---|
| Local | Conventional formula | Detailed | Directly yield | Equation (10) |
| | Matrix manipulation | Simple | Directly yield | Equation (9) |
| | Spatial correlation matrix | Simple | Directly yield | Equations (15) and (16) |
| | Potential energy | Moderate | Indirectly yield | Equations (47)-(50) |
| Global | Conventional formula | Detailed | Directly yield | Equation (8) |
| | Three-step calculation | Very simple | Directly yield | Equations (3), (5), and (7) |
| | Matrix scaling | Simple | Directly yield | Equation (13) or (26) |
| | Linear regression | Moderate | Directly yield | Equation (33) or (34) |
| | Local weighting | Moderate | Indirectly yield | Equation (40) |
| | Spatial correlation matrix | Simple | Indirectly yield | Equations (15) and (16) |
| | Outer product sum | Simple | Directly yield | Equations (33) and (54) |

**Note**: If the utilized variable $y$ is replaced by the standardized variable $z$, the seven approaches can be employed to evaluate global Moran's $I$, for which the seventh method can also be termed standard deviation method.

## 4.2 Empirical analysis

The new framework of spatial autocorrelation based on Getis-Ord's indexes can be applied to China's cities to make case studies. The study area includes the whole mainland of China, and the time points are 2000 and 2010, respectively. As an example of illustrating a methodology, the simpler, the better. Therefore, only the capital cities of the 31 provinces, autonomous regions, and municipalities directly under the Central Government of China (CCC) are taken into account. The urban population from the fifth census in 2000 and the sixth census in 2010 can serve as the two size variables ($x_i$), and the railway mileage between any two cities are used as a spatial proximity measurement ($r_{ij}$). Because the cities of Haikou and Lhasa were not connected to Chinese network of cities by railway for a long time, only 29 cities are really considered in the spatial analysis, and thus the size of the spatial sample is $n$=29.

Using the methods shown above and the datasets of city sizes and spatial distances, we can calculate the Getis-Ord's indexes and potential energy measurements of Chinese systems of cities. By means of one of the seven approaches above-shown, we can compute the global Getis-Ord's index. For example, using the three-step method based on the formula $G=\mathbf{y}^{\mathrm{T}}\mathbf{W}\mathbf{y}$, we have the following results, for 2000 year, $G$=0.001299, and for 2010 year, $G$=0.001345. By using one of the four approaches displayed above, we can compute the local Getis-Ord's indexes. On the other, using the formula of potential energy index and mutual energy index ($K$=1, $q$=1), equations (43) and (44), we can compute the potential energy indexes and mutual energy indexes. If $K$=1 and $q$=1 as given, then the potential energy indexes equal the corresponding the local Getis-Ord's indexes, and the mutual energy indexes are just the product of unitized size variable and the local Getis-Ord's indexes. In short, local Getis-Ord's indexes equal the normalized potential energy indexes, and the sum of the mutual energy indexes equals the global Getis-Ord's index (Table 3).

**Table 3 The main computational results of spatial autocorrelation and spatial interaction based on Getis-Ord's indexes (2000 & 2010)**

| City | 2000 | | | 2010 | | |
|---|---|---|---|---|---|---|
| | Variable ($y_i$) | Local $G_i$ & PEI ($V_i$) | $yG_i$ & MEI ($E_i$) | Variable ($y_i$) | Local $G_i$ & PEI ($V_i$) | $yG_i$ & MEI ($E_i$) |
| **Beijing** | 0.096014 | 0.001774 | 0.000170 | 0.109598 | 0.001831 | 0.000201 |
| **Changchun** | 0.027262 | 0.001172 | 0.000032 | 0.023185 | 0.001162 | 0.000027 |
| **Changsha** | 0.021463 | 0.001403 | 0.000030 | 0.020274 | 0.001346 | 0.000027 |
| **Chengdu** | 0.038637 | 0.000938 | 0.000036 | 0.041530 | 0.000938 | 0.000039 |
| **Chongqing** | 0.057390 | 0.000907 | 0.000052 | 0.061105 | 0.000898 | 0.000055 |
| **Fuzhou** | 0.020029 | 0.000925 | 0.000019 | 0.018852 | 0.000915 | 0.000017 |
| **Guangzhou** | 0.069445 | 0.000784 | 0.000054 | 0.065137 | 0.000776 | 0.000051 |
| **Guiyang** | 0.018497 | 0.001008 | 0.000019 | 0.017128 | 0.001009 | 0.000017 |
| **Hangzhou** | 0.024784 | 0.001985 | 0.000049 | 0.031087 | 0.001969 | 0.000061 |
| **Harbin** | 0.034932 | 0.000931 | 0.000033 | 0.032845 | 0.000911 | 0.000030 |
| **Hefei** | 0.014790 | 0.001580 | 0.000023 | 0.021679 | 0.001594 | 0.000035 |
| **Hohhot** | 0.010019 | 0.001082 | 0.000011 | 0.010124 | 0.001106 | 0.000011 |
| **Jinan** | 0.026145 | 0.001690 | 0.000044 | 0.023697 | 0.001751 | 0.000042 |
| **Kunming** | 0.025059 | 0.000705 | 0.000018 | 0.022152 | 0.000704 | 0.000016 |
| **Lanzhou** | 0.018354 | 0.000931 | 0.000017 | 0.016780 | 0.000934 | 0.000016 |
| **Nanchang** | 0.016881 | 0.001512 | 0.000026 | 0.013512 | 0.001490 | 0.000020 |
| **Nanjing** | 0.034852 | 0.001766 | 0.000062 | 0.039725 | 0.001785 | 0.000071 |
| **Nanning** | 0.013695 | 0.000812 | 0.000011 | 0.017085 | 0.000798 | 0.000014 |
| **Shanghai** | 0.128610 | 0.001205 | 0.000155 | 0.124315 | 0.001278 | 0.000159 |
| **Shenyang** | 0.043929 | 0.001130 | 0.000050 | 0.039929 | 0.001139 | 0.000045 |
| **Shijiazhuang** | 0.019519 | 0.002036 | 0.000040 | 0.019428 | 0.002084 | 0.000040 |
| **Taiyuan** | 0.025663 | 0.001529 | 0.000039 | 0.021558 | 0.001565 | 0.000034 |
| **Tianjin** | 0.053723 | 0.002228 | 0.000120 | 0.062410 | 0.002345 | 0.000146 |
| **Urumqi** | 0.017468 | 0.000420 | 0.000007 | 0.019647 | 0.000420 | 0.000008 |
| **Wuhan** | 0.066318 | 0.001269 | 0.000084 | 0.051300 | 0.001277 | 0.000066 |
| **Xi'an** | 0.036855 | 0.001200 | 0.000044 | 0.034418 | 0.001204 | 0.000041 |
| **Xining** | 0.008639 | 0.000890 | 0.000008 | 0.008041 | 0.000883 | 0.000007 |
| **Yinchuan** | 0.005847 | 0.000938 | 0.000005 | 0.007895 | 0.000937 | 0.000007 |
| **Zhengzhou** | 0.025183 | 0.001665 | 0.000042 | 0.025565 | 0.001660 | 0.000042 |
| **Sum** | **1.000000** | **0.036414** | **0.001299** | **1.000000** | **0.036710** | **0.001345** |
| **Mean** | **0.034483** | **0.001256** | **0.000045** | **0.034483** | **0.001266** | **0.000046** |

**Note**: The sum of the $E_i$ values is equal to the global Getis-Ord's index.

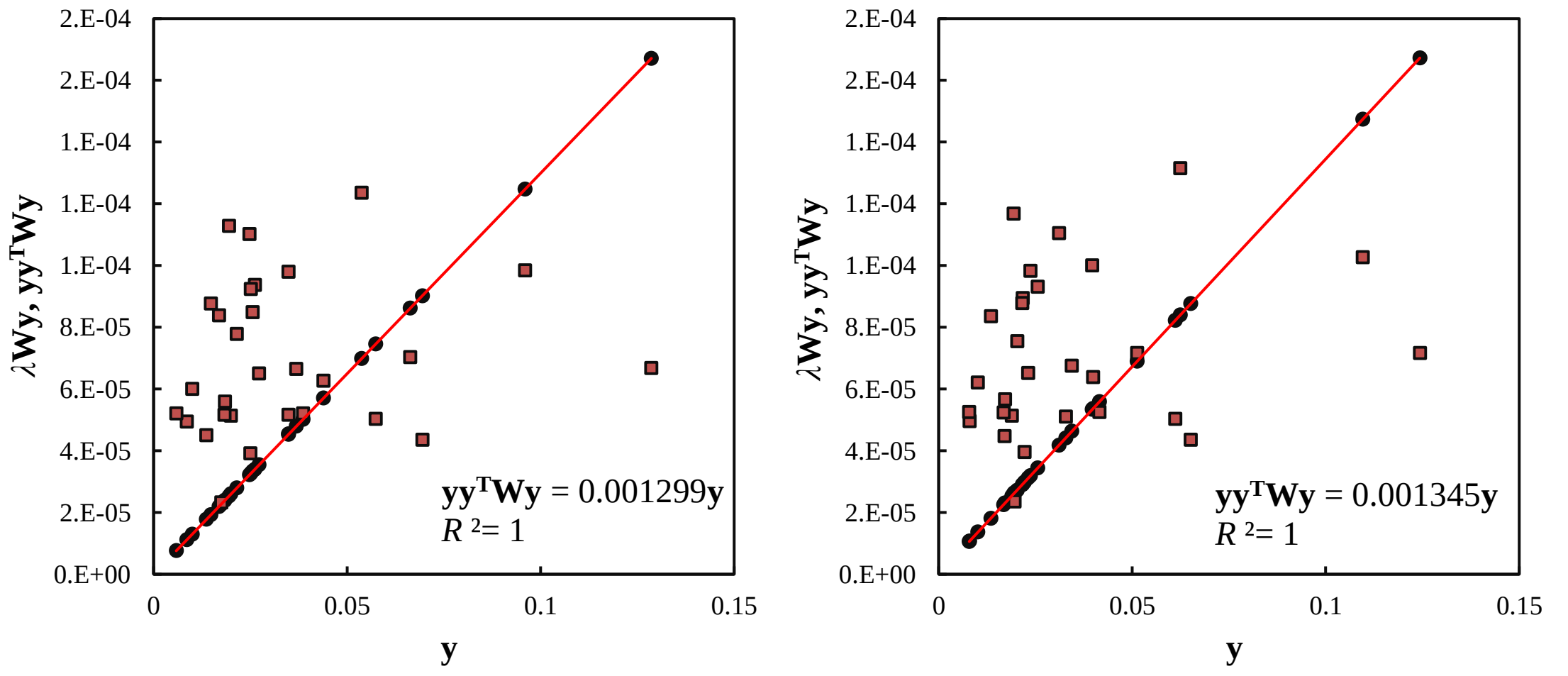

(a) 2000 (b) 2010

**Figure 2 The scatterplots of spatial auto-correlation based on Getis-Ord's measurement for the main cities of China (2000 & 2010)**

(**Note**: The trend line is added to the trend points based on the outer product correlation, $\mathbf{yy}^T\mathbf{Wy}$, and we have perfect fit, $R^2$=1. This implies that the connection line of the scattered points yielded by the linear relation between $\mathbf{y}$ and $\mathbf{yy}^T\mathbf{Wy}$ is just the trend line. )

Furthermore, we can draw the Getis-Ord's scatterplots by means of the scaling relation between the unitized size vectors and the spatial correlation matrixes. Using equations (33) and (34), we have two variables $f=\lambda\mathbf{Wy}$ and $f^{*}=\mathbf{yy}^{\mathrm{T}}\mathbf{Wy}$ (Table 4). The relationships between $y$ and $f(y)$ give a scatter plot, and relationships between $y$ and $f^{*}(y)$ yields a trend line in the scatter plot (Figure 2). The scatter plot has at least three uses. First, it can be used to estimate the global Getis-Ord's index. The slope of the trend line is equal to global Getis-Ord's $G$. Second, it can be used to reflect the spatial distribution feature of a geographical system. Third, it can be used to make a simple classification for the research objects. If the points are above the trend line, the actual values of the potential energy indexes are greater than the expected values; if the points are below the trend line, the actual potential energy index values are less than the expected values. Specially, if the points are on the trend line, the actual values are close to the expected values of the potential energy indexes. A discriminant index for the simple classification can be defined as

$$h_i = \frac{f_i}{f_i^{*}} = \frac{(\mathbf{y}^{\mathrm{T}}\mathbf{y}\mathbf{W}\mathbf{y})_i}{(\mathbf{y}\mathbf{y}^{\mathrm{T}}\mathbf{W}\mathbf{y})_i}, \tag{55}$$

where $h_i$ denotes the discriminant index. If $h_i>1$, the $i$th point is above the trend line, otherwise, the point is beneath the trend line. By the way, the trend line represents the conditional mean value, and the potential energy indexes are equal to the local Getis-Ord's indexes and indicate accessibility.

**Table 4 The computational results of spatial autocorrelation for Getis-Ord's scattered plots (2000 & 2010)**

| City | 2000 | | | 2010 | | |
|---|---|---|---|---|---|---|
| | Variable ($y$) | $\mathbf{y}^{\mathrm{T}}\mathbf{yWy}$ ($f$) | $\mathbf{yy}^{\mathrm{T}}\mathbf{Wy}$ ($f^{*}$) | Variable ($y$) | $\mathbf{y}^{\mathrm{T}}\mathbf{yWy}$ ($f$) | $\mathbf{yy}^{\mathrm{T}}\mathbf{Wy}$ ($f^{*}$) |
| **Beijing** | 0.096014 | 0.000098 | 0.000125 | 0.109598 | 0.000103 | 0.000147 |
| **Changchun** | 0.027262 | 0.000065 | 0.000035 | 0.023185 | 0.000065 | 0.000031 |
| **Changsha** | 0.021463 | 0.000078 | 0.000028 | 0.020274 | 0.000076 | 0.000027 |
| **Chengdu** | 0.038637 | 0.000052 | 0.000050 | 0.041530 | 0.000053 | 0.000056 |
| **Chongqing** | 0.057390 | 0.000050 | 0.000075 | 0.061105 | 0.000050 | 0.000082 |
| **Fuzhou** | 0.020029 | 0.000051 | 0.000026 | 0.018852 | 0.000051 | 0.000025 |
| **Guangzhou** | 0.069445 | 0.000044 | 0.000090 | 0.065137 | 0.000044 | 0.000088 |
| **Guiyang** | 0.018497 | 0.000056 | 0.000024 | 0.017128 | 0.000057 | 0.000023 |
| **Hangzhou** | 0.024784 | 0.000110 | 0.000032 | 0.031087 | 0.000110 | 0.000042 |
| **Harbin** | 0.034932 | 0.000052 | 0.000045 | 0.032845 | 0.000051 | 0.000044 |
| **Hefei** | 0.014790 | 0.000088 | 0.000019 | 0.021679 | 0.000089 | 0.000029 |
| **Hohhot** | 0.010019 | 0.000060 | 0.000013 | 0.010124 | 0.000062 | 0.000014 |
| **Jinan** | 0.026145 | 0.000094 | 0.000034 | 0.023697 | 0.000098 | 0.000032 |
| **Kunming** | 0.025059 | 0.000039 | 0.000033 | 0.022152 | 0.000040 | 0.000030 |
| **Lanzhou** | 0.018354 | 0.000052 | 0.000024 | 0.016780 | 0.000052 | 0.000023 |
| **Nanchang** | 0.016881 | 0.000084 | 0.000022 | 0.013512 | 0.000084 | 0.000018 |
| **Nanjing** | 0.034852 | 0.000098 | 0.000045 | 0.039725 | 0.000100 | 0.000053 |
| **Nanning** | 0.013695 | 0.000045 | 0.000018 | 0.017085 | 0.000045 | 0.000023 |
| **Shanghai** | 0.128610 | 0.000067 | 0.000167 | 0.124315 | 0.000072 | 0.000167 |
| **Shenyang** | 0.043929 | 0.000063 | 0.000057 | 0.039929 | 0.000064 | 0.000054 |
| **Shijiazhuang** | 0.019519 | 0.000113 | 0.000025 | 0.019428 | 0.000117 | 0.000026 |
| **Taiyuan** | 0.025663 | 0.000085 | 0.000033 | 0.021558 | 0.000088 | 0.000029 |
| **Tianjin** | 0.053723 | 0.000124 | 0.000070 | 0.062410 | 0.000132 | 0.000084 |
| **Urumqi** | 0.017468 | 0.000023 | 0.000023 | 0.019647 | 0.000024 | 0.000026 |
| **Wuhan** | 0.066318 | 0.000070 | 0.000086 | 0.051300 | 0.000072 | 0.000069 |
| **Xi'an** | 0.036855 | 0.000067 | 0.000048 | 0.034418 | 0.000068 | 0.000046 |
| **Xining** | 0.008639 | 0.000049 | 0.000011 | 0.008041 | 0.000050 | 0.000011 |
| **Yinchuan** | 0.005847 | 0.000052 | 0.000008 | 0.007895 | 0.000053 | 0.000011 |
| **Zhengzhou** | 0.025183 | 0.000092 | 0.000033 | 0.025565 | 0.000093 | 0.000034 |
| **Sum** | **1.000000** | **0.002020** | **0.001299** | **1.000000** | **0.002059** | **0.001345** |
| **Mean** | **0.034483** | **0.000070** | **0.000045** | **0.034483** | **0.000071** | **0.000046** |

**Note**: The sum of the $f_i^*$ values is equal to the global Getis-Ord's index.

About the Getis-Ord's scatter plot, it is necessary to explain the two aspects. First, generally speaking, the scattered points are not consistent with the trend line. If we fit equation (34) to the dataset based on the relationship between $\lambda\mathbf{Wy}$ and $\mathbf{y}$, the slope of the trend line gives the regression coefficient, which represents the expected global Getis-Ord's index. Second, there is an alternative form for the scatter plot. If we substitute the original $x$-axis represented by $y$ with $\mathbf{f}^*=\mathbf{yy}^T\mathbf{Wy}$, the pattern of the scattered points have no change. In other words, we can use the relationships between $f^*$ and $f$ to replace the relationships between $y$ and $f$ (Figure 4). The relative spatial relationships between the scattered points do not change despite the variable substitution. The difference is that the trend line is superseded by the diagonal line from the lower left corner to the upper right corner ($f^*=f$). The scatterplots show that 5 or 6 points are prominent. In 2000, five points are significantly below the trend lines, and these points represent Beijing, Chongqing, Guangzhou, Shanghai, and Wuhan; in 2010, six cities are significantly below the trend line, that is, Beijing, Chongqing, Guangzhou, Shanghai, Chengdu, and Urumqi. Among these cities below the trend line, three ones are the municipalities directly under CCC: Beijing, Chongqing, and Shanghai. Among the four municipalities directly under CCC, Tianjin is a special case or exception. The point representing Tianjin is significantly above the trend line, indicating the highest potential energy index.

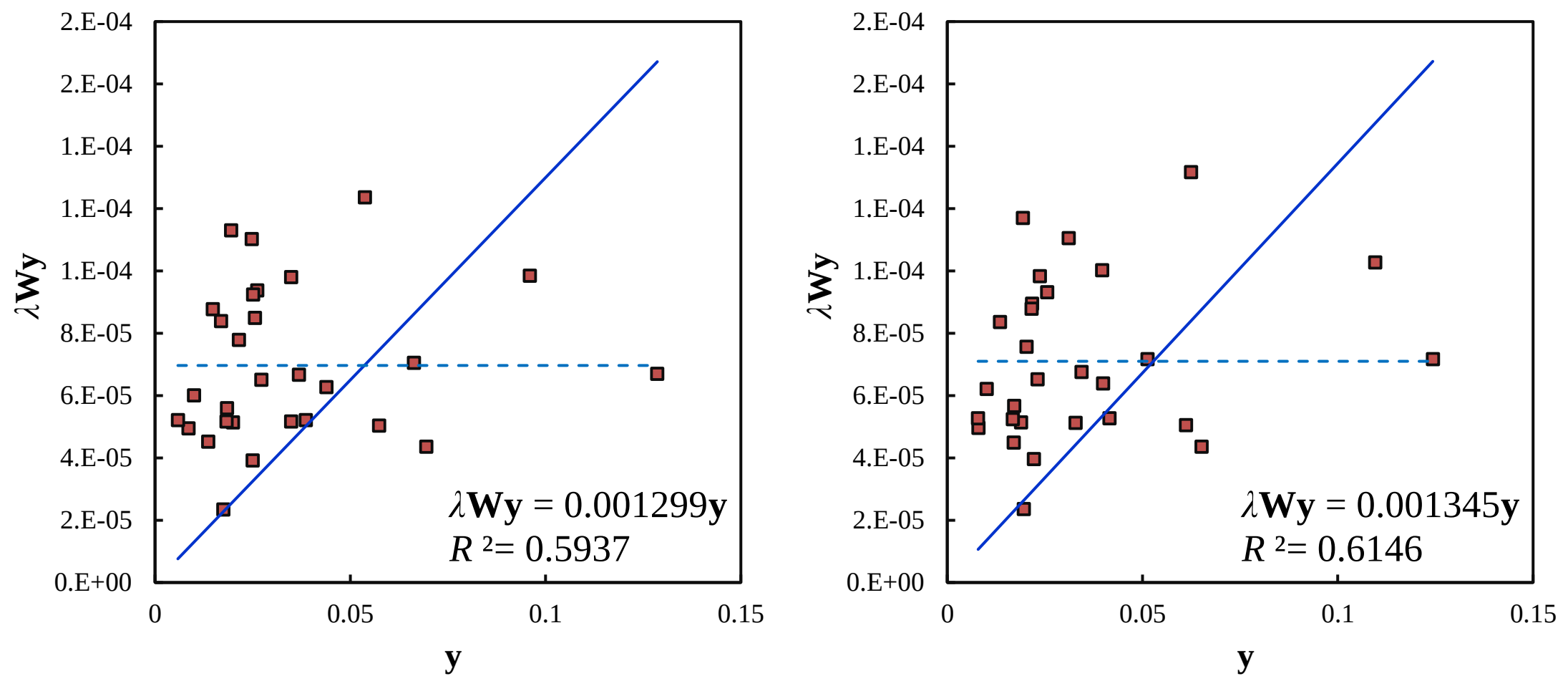

(a) 2000 (b) 2010

**Figure 3 The normal parameter values and abnormal goodness of fit in the scatterplots of spatial auto-correlation based on Getis-Ord's indexes for the main cities of China (2000 & 2010)**

(**Note**: The trend line is added to the scattered points based on inner product correlation, $\lambda$**Wy**, and the intercept is set as 0. The slope of the trend line give the global Getis-Ord's index, and the value of goodness of fit, $R^2$, is defined by cosine instead of Pearson correlation. The horizontal line represent absolute average line.)

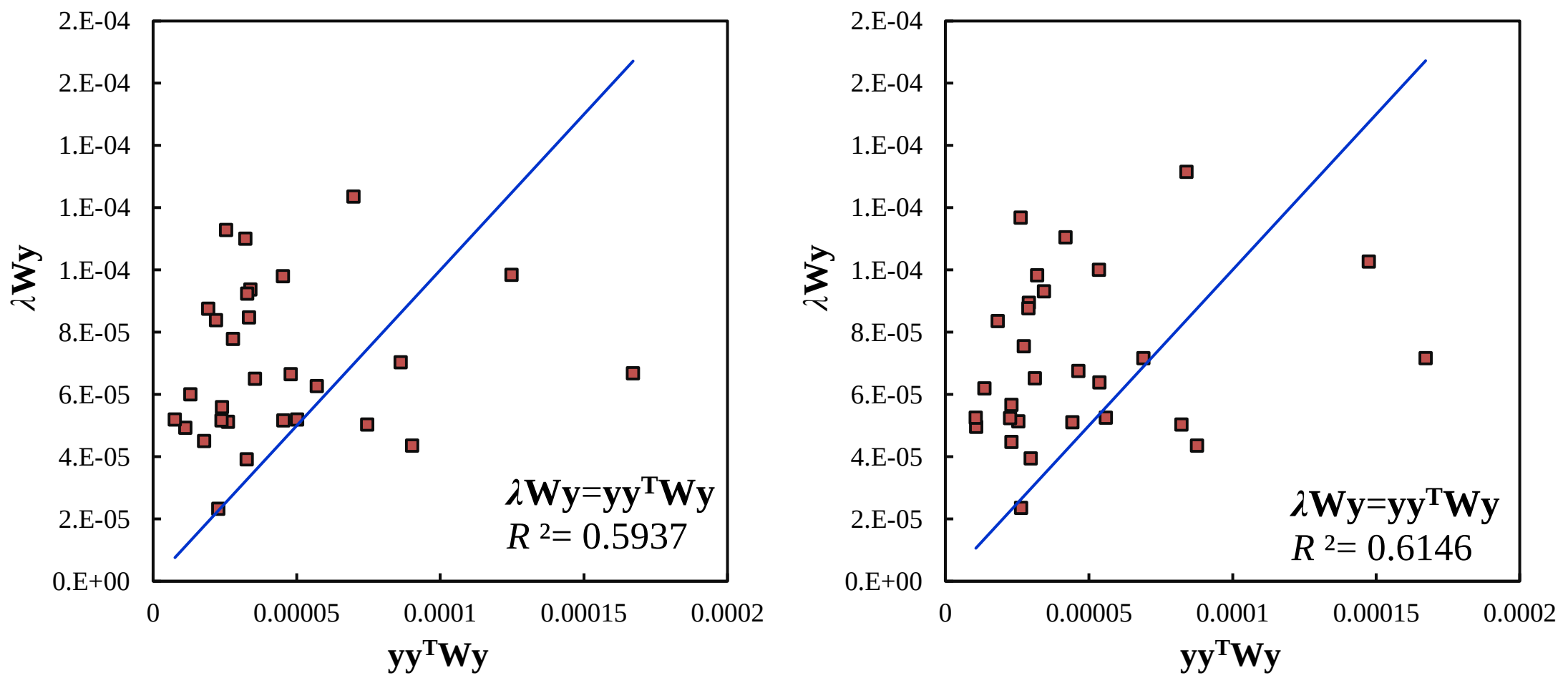

(a) 2010 (b) 2010

**Figure 4 The alternative forms of the scatterplots of spatial auto-correlation based on Getis-Ord's measurement for the main cities of China (2010 & 2010)**

(**Note**: This scatter plot is equivalent to the ones display in Figure 3, but the variable **y** used as a horizontal axis is replaced by the new variable $\mathbf{f}^{*}=\mathbf{yy}^{\mathrm{T}}\mathbf{Wy}$. In this case, the original trend line is replaced by a diagonal line.)

The abovementioned trend line represents conditional mean. Moreover, the arithmetic mean result represents the absolute mean. The absolute mean forms a horizontal average line. The scatterplot can be divided into four “quadrants” by using the conditional mean and absolute mean. In 2000, the absolute mean of the potential energy indexes is about 0.000070, and in 2010, the absolute mean is around 0.000071. If we add the average line reflecting absolute means to a Getis-Ord’s scatterplot, the 29 main cities of China will fall into four sub-regions. The meanings of the four sub-region are as follows. (I) The first region is the upper right part, representing high-high type quadrant (H-H type). The potential energy index of a city is high, so are the potential indexes of surrounding cities. The typical city is Beijing, the national capital of China. (II) The second region is the upper left part, representing high-low type quadrant (H-L type). The potential energy index of a city is high, and there are cities with low potential indexes around it. The typical cities are Tianjin and Hangzhou. (III) The third region is the lower left part, representing the low-low type quadrant (L-L type). The potential energy index of a city is low, and there are cities with low potential indexes around it. The typical cities are Kunming and Nanning. (V) The fourth region is the lower right part, representing the low-high quadrant (L-H type). The potential energy index of a city is low, and there are cities with high potential index around it. The typical cities are Chongqing and Guangzhou. Of course, the high and low potential energy indexes are relative to one another. From 2000 to 2010, only Shanghai, Wuhan, Chengdu, and Urumqi have changed their situations. In fact, Chengdu and Urumqi are near the trend line, their *h* values are close to 1. This means that their category characteristics are not obvious. Nevertheless, this classification outlines a clear map of urban location and spatial correlation of cities in Mainland China (Table 5).

**Table 5 Chinese city classification based on conditional mean (trend line) and absolute mean (average line) (2000 & 2010)**

| Quadrant | 2000 | 2010 |
|---|---|---|
| **I (H-H)** | Beijing, Wuhan | Beijing, Shanghai |
| **II (H-L)** | Tianjin, Shijiazhuang, Hangzhou, Nanjing, Jinan, Zhengzhou, Hefei, Taiyuan, Nanchang, Changsha | Tianjin, Shijiazhuang, Hangzhou, Nanjing, Jinan, Zhengzhou, Hefei, Taiyuan, Nanchang, Changsha, Wuhan |
| **III (L-L)** | Xi'an, Changchun, Shenyang, Hohhot, Guiyang, Chengdu, Yinchuan, Lanzhou, Harbin, Fuzhou, Xining, Nanning, Kunming, Urumqi | Xi'an, Changchun, Shenyang, Hohhot, Guiyang, Yinchuan, Lanzhou, Harbin, Fuzhou, Xining, Nanning, Kunming |
| **V (L-H)** | Shanghai, Chongqing, Guangzhou | Chongqing, Guangzhou, Chengdu, Urumqi |

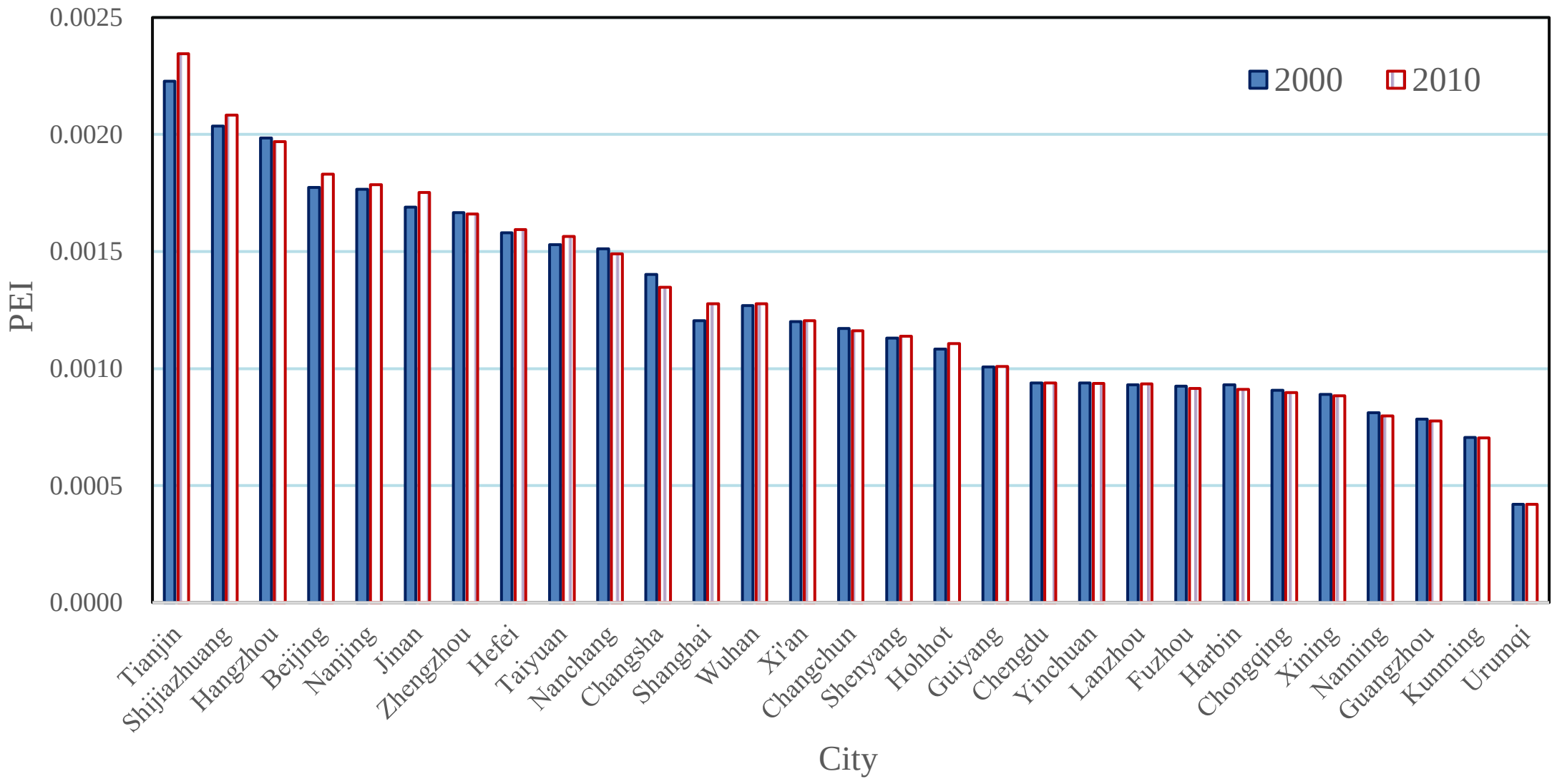

**Figure 5 The potential energy indexes and local Getis-Ord's indexes of the main cities in Mainland China (2000 & 2010)**

The locational properties and the spatial association of the 29 Chinese cities can be evaluated by the potential energy indexes and mutual energy indexes. The local Getis-Ord's indexes are equivalent to the normalized potential energy indexes, and the sum of the mutual energy index equals the global Getis-Ord's index. By way of potential and mutual energy concepts, we can understand Getis-Ord's statistics deeply. Using local Getis-Ord's indexes or potential energy indexes of Chinese cities, we can evaluate the traffic accessibility of these cities. The main features are as follows. First, if the size of a city is relatively small, but there is big cities near the city, then its potential index is high. The typical cities are Tianjin, Shijiazhuang, Hangzhou, and Nanjing. Tianjin and Shijiazhuang are adjacent to the megacity, Beijing, while Hangzhou and Nanjing are adjacent to the megacity, Shanghai. Second, if a city is in the center of the network of cities, then its potential energy index is relatively high to some extent. The typical city is Zhengzhou. The location of Wuhan is also superior, but its size is too large to increase its potential index. Third, the cities in remote areas bear lower potential indexes due to being far from the city network of Chinese mainland. The typical city is Urumqi in Xinjiang, northwestern China, having the lowest potential index. The next one to last is Kunming in Yunnan, located in southwestern China. Although Guangzhou is an economically developed city, due to its location on the southern sea coast, its potential index is also in the bottom. Fourth, during the period from 2000 to 2010, the potential energy indexes of these cities have no significant change. This suggests that the potential indexes of the main Chinese cities are very stable (Figure 5). An interesting phenomenon is that because there are no other large cities around Urumqi, it turned into a high-low type of city in 2000.

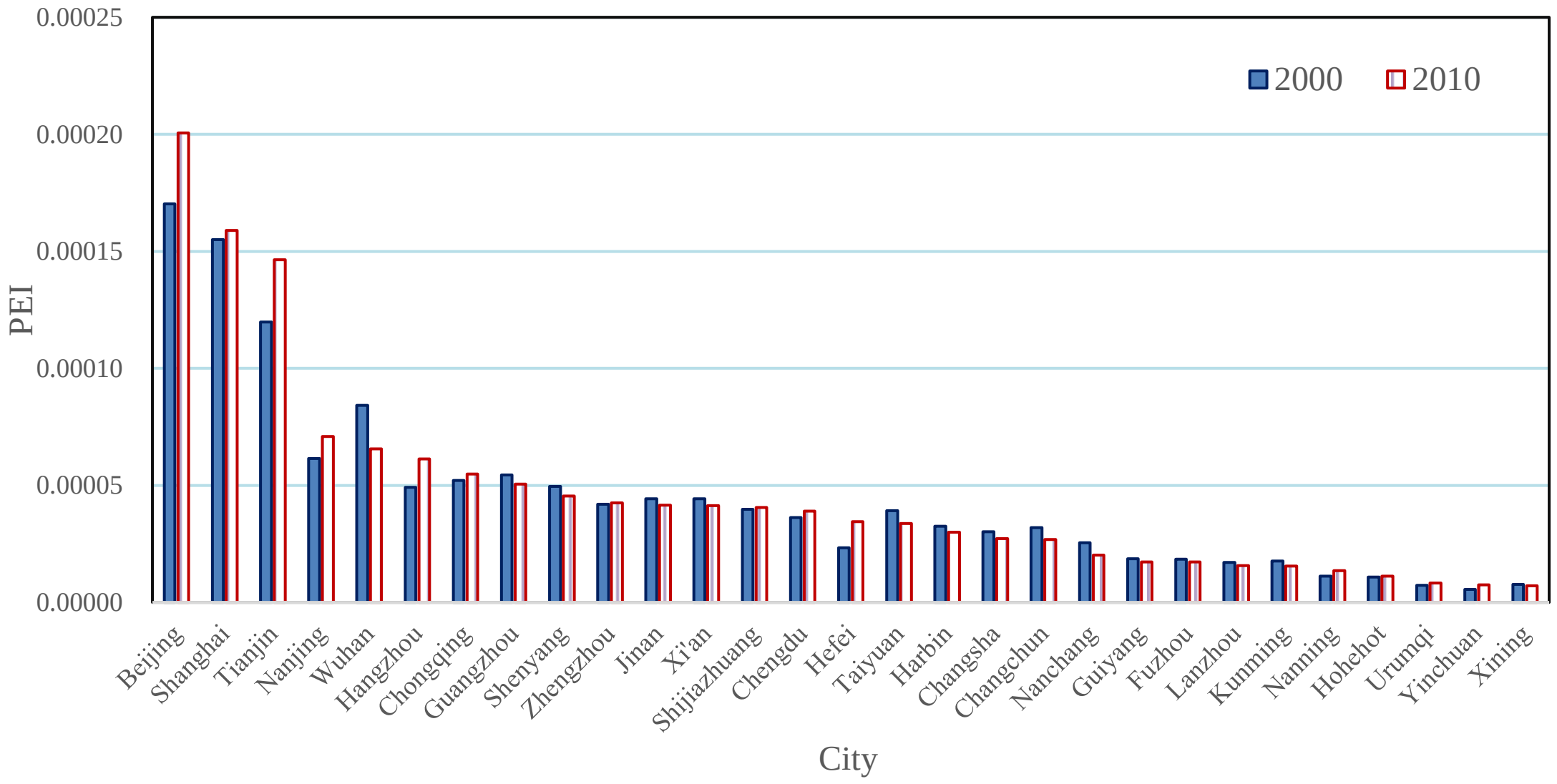

**Figure 6 The mutual energy indexes based on census population of the main cities in Mainland China (2000 & 2010)**

The potential energy index depends on the location of a city in an urban network, but it has nothing to do directly with the size of the city itself. So the potential energy indexes and thus local Getis-Ord's indexes reflect the spatial association rather than spatial influence. Reflecting the influence power of a city in a network of cities, the mutual energy indexes are function of city size and potential energy indexes. As indicated above, the potential energy index implies a city's accessibility of transportation and the superiority of geographical location in an urban network. Using the mutual energy indexes of the 29 Chinese cities, we can illustrate the absolute positions of these cities in the urban network (Figure 6). The top cities of spatial influence are Beijing, Shanghai, and Tianjin, which are the old municipalities directly under the Central Government of China. From 2000 to 2010, the mutual energy indexes of the three municipalities have significant change. After the three old municipalities, the cities with higher mutual energy index values include Nanjing, Hangzhou, Wuhan, Hangzhou, and Chongqing, which have superior geographic locations and large city sizes. The cities in marginal areas, such as Xining, Yinchuan, Hohhot, Nanning, Kunming, Lanzhou, Fuzhou, and Guiyang bear lower mutual energy indexes due to small city sizes and geographical locations away from the center of urban network. The cities like Xi'an, Shijiazhuang, Chengdu, Harbin, and so on, have middle mutual energy indexes owing to one of advantages in city size or geographical location. The mutual energy index of Hefei went up fast because of city population size doubled from 2000 to 2010.

## 5 Conclusions

Scientific research involves two elements, that is, description and understanding. Getis-Ord's indexes are a type of statistic measurements for spatial description. So, geographical explanation is not the main aim of this study. As a work of methodology research, this paper is devoted to normalizing, developing, and improving the analytical process and techniques of the spatial autocorrelation modeling based on Getis-Ord's indexes. The chief contributions of this work to geographical spatial analysis lie in four aspects: (1) the computational process is significantly simplified and diversified, (2) the scatter plot is introduced into the analytical process, (3) the parameter characters of the global and local Getis-Ord's indexes are illustrated, and (4), the relationship between Getis-Ord's index and potential energy is revealed. If the spatial contiguity matrix is generated using power-law decay function, the local Getis-Ord's indexes proved to be equivalent to potential energy measurements. Based on these results and findings, we can reach the main conclusions as follows. **First, the prerequisite for the effective use of Getis-Ord's indexes is that the spatial distributions and size distribution possess characteristic scales.** The global Getis-Ord's index, which is a weighted sum of local indexes, is an eigenvalue of spatial correlation matrix, and the local indexes form an eigenvector of the outer product matrix of the unitized size vector. This suggests that the global index is a characteristic length of spatial correlation. For the scale-free geographical processes and patterns, the Getis-Ord's index is no longer valid. What is more, the unitization processing of size variable depends the average value, where represents the characteristic length of statistical analysis. This implies that we need new measurement for scale-free spatial autocorrelation. **Second, the spatial autocorrelation and spatial interaction can be integrated into an analytical framework.** The Getis-Ord's indexes are the measurements for spatial autocorrelation, while the potential energy indexes are the measurement based on spatial interaction. However, the two kinds of measurements are equivalent to one another if the distance decay function is an inverse power law. By unitizing size vector and rescaling spatial distances, we can obtain Getis-Ord's indexes by calculating potential energy indexes. This indicates that we can unify spatial autocorrelation and spatial interaction to a degree by means of spatial correlation functions. **Third, the spatial analytical processes based on Getis-Ord's indexes can be visualized by normalized scatterplot.** The scatterplot similar to Moran's plot can be employed to make both spatial autocorrelation and spatial interaction analyses in the new framework. The scatterplot can provide a visual pattern for spatial modeling results. Using the scattered points indicating observational values, the trend line indicating predicted values, and the average line indicating absolute mean of local potential energy indexes, we can make a simple spatial cluster for geographical elements in a study area. In practice, different researchers may obtain different types of geographical information from the scatter plots and the related cluster results.

## Acknowledgement:

This research was sponsored by the National Natural Science Foundation of China (Grant No. 41671167. See: http://isisn.nsfc.gov.cn/egrantweb/). The support is gratefully acknowledged.